\documentclass[12pt,preprint]{emulateapj}

\newcommand{\myemail}{redelson@astro.umd.edu}
\newcommand{\zw}{Zw~229$-$15}
\newcommand{\kepler}{{\it Kepler}}

\slugcomment{To appear in the Astrophysical Journal}

\shorttitle{Power spectrum of \zw}
\shortauthors{R. Edelson et al.}

\begin{document}

\title{Discovery of a $\sim 5$ day characteristic timescale in the Kepler power spectrum of Zw 229-15}

\author{R. Edelson}
\affil{Department of Astronomy, University of Maryland, College Park, MD 20742-2421, USA}
\email{\myemail}

\author{S. Vaughan}
\affil{University of Leicester, X-ray and Observational Astronomy Group, Department of Physics and Astronomy, University Road, Leicester LE1 7RH, UK}

\author{M. Malkan}
\affil{Department of Physics \& Astronomy, University of California, Los Angeles, CA 90095-1547, USA}

\author{B. C. Kelly}
\affil{Department of Physics, Broida Hall, University of California, Santa Barbara, CA 93106-9530, USA}

\author{K. L. Smith}
\affil{Department of Astronomy, University of Maryland, College Park, MD 20742-2421, USA}

\author{P. T. Boyd}
\affil{Astrophysics Science Division, NASA/GSFC, Code 660, Greenbelt, MD 20771, USA}

\author{R. Mushotzky}
\affil{Department of Astronomy, University of Maryland, College Park, MD 20742-2421, USA}

\begin{abstract}
We present time series analyses of the full \kepler\ dataset of \zw. 
This \kepler\ light curve --- with a baseline greater than three years, composed of virtually continuous, evenly sampled 30-minute measurements --- is unprecedented in its quality and precision.
We utilize two methods of power spectral analysis to investigate the optical variability and search for evidence of a bend frequency associated with a characteristic optical variability timescale. 
Each method yields similar results.
The first interpolates across data gaps to use the standard Fourier periodogram.
The second, using the CARMA-based time-domain modeling technique of \citet{Kelly14}, does not need evenly-sampled data. 
Both methods find excess power at high frequencies that may be due to \kepler\ instrumental effects.
More importantly both also show strong bends ($\Delta \alpha \sim 2$) at timescales of $\sim 5$ days, a feature similar to those seen in the X-ray PSDs of AGN but never before in the optical.
This observed $\sim 5$ day timescale may be associated with one of several physical processes potentially responsible for the variability.
A plausible association could be made with light-crossing, dynamical or thermal timescales, depending on the assumed value of the accretion disk size and on unobserved disk parameters such as $\alpha$ and $H/R$.
This timescale is not consistent with the viscous timescale, which would be years in a $\sim 10^7 M_\odot$ AGN such as \zw.
However there must be a second bend on long ($\gtrsim 1$ year) timescales, and that feature could be associated with the viscous timescale.
\end{abstract}


\section{Introduction}
\label{sect:intro}

The accretion disks of Active Galactic Nuclei (AGN) are much too distant to image directly.
Simple physical arguments place an upper limit on the accretion disk size at $\sim 10^{-2}$ parsecs \citep{Hawkins07}, corresponding to $\lesssim 1$ milliarcsec for even the closest AGN.
Thus indirect methods must be used to study their centers.  
AGN's strong and rapid aperiodic optical variability provides a powerful tool for constraining physical conditions and processes.  
For example reverberation mapping allows estimation of Seyfert 1 emission-line region sizes and central black hole masses \citep{Peterson04}.  
However recent progress has been slowed by limitations inherent in ground-based optical monitoring where it is nearly impossible to obtain continuous light curves longer than $\sim 12$ hr and errors better than $\sim 1$\%.

The \kepler\ mission breaks these barriers with fast (30 min) sampling, high ($>90$\%) duty cycle, and excellent ($\sim 0.1$\% for a $16$th magnitude source) precision.  
\zw\ ($z = 0.025$; \citealt{Falco99}) was observed during Quarters 4-17 (Q4-17), totaling $3.4$ years.  
As the longest-monitored, brightest, and one of the most strongly variable \kepler\ AGN, it is ideal for time-series analyses that estimate the Power Spectral Density 
(PSD) function.
Previous studies used only a small fraction of these data, finding a power law-like PSD with a steep index ($P_f \propto f^\alpha$; \citealt{Mushotzky11, Kelly14}) with indications of a break on $\sim 90$ day timescales \citep{Carini12}.

This paper reports light curve extraction and PSD analysis of the full \kepler\ dataset.
The paper is organized as follows: Section \ref{sect:obs} reports the data reduction, Section \ref{sect:psd} presents the PSD analyses, Section \ref{sect:disco} discusses the theoretical implications of these analyses, and Section \ref{sect:concl} concludes with a brief  summary of this work and future plans.

\section{Data reduction}
\label{sect:obs}

Since \kepler\ was designed to detect exoplanets, the standard pipeline processing removes long-term trends from light curves to optimize detection of short, shallow dips.
This renders data unusable for AGN, which show broad intrinsic variability power over timescales of hours to years.  
Thus previous \kepler\ \zw\ studies (e.g., \citealt{Mushotzky11}) used ``Simple Aperture Photometry'' (SAP) data from earlier in the pipeline, but this also contained uncorrected systematics due to differently-sized ``optimal'' extraction apertures.

\subsection{Extraction from pixel data}

In order to avoid these systematics, our approach starts earlier with the 2-dimensional calibrated pixel data.  
We used the PyKE\footnote{Software available at {\tt http://keplergo.arc.nasa.gov/PyKE.shtml}} programs {\tt kepmask} and {\tt kepextract} to build large 32-pixel masks (see Figure 1). 
These larger masks mitigate excursions due to thermally-induced focus changes and to differential velocity aberration (\cite{Kinemuchi12}; see also {\tt http://keplergo.arc.nasa.gov/PyKEprimerTPFs.shtml}).
These extraction masks are identical within each season (e.g. Season 2 = Q4/8/12/16) even though the pixel downloads are not.

The resulting 3.4 year (55,653 cadence) light curve is shown in the top panel of Figure 2.
The interquarter jumps arise because quarterly spacecraft rolls move the source to a different chip with a different aperture.  
Thanks to \kepler's stable pointing and these identical seasonal masks, the jumps between the same seasons are highly repeatable.

\subsection{Interquarter scaling}

Previous \kepler\ AGN studies dealt with interquarter jumps by restricting the analyses to single quarters \citep{Mushotzky11, Kelly14}, performing simple scaling to match fluxes across the gaps \citep{Wehrle13} or using ground-based data to normalize the offsets \citep{Carini12}.  
Our approach starts by averaging the five good cadences immediately before and after each gap and taking the difference, then sorting these by season (Table 1). 
We then measured and applied mean seasonal corrections (e.g., the mean difference between end/start of quarters 4/5, 8/9, 12/13, 16/17).  
A good measure of the systematics remaining after these corrections is the standard deviation of the residuals: $63$ ct s$^{-1}$.  
Since the mean flux was $15,933$ ct s$^{-1}$, this indicates $\sim 0.4$\% residual errors after quarterly offset correction.

\subsection{Filtering}

Next, bad cadences are filtered on three criteria: 1) ``manual exclude'' ({\tt SAP\_QUALITY} bit 9 set true), indicating Solar coronal mass ejections, 2) four cadence ranges ($54,941$-$54,960$, $64,084$-$64,129$, $71,054$-$71,063$, $72,324$-$72,332$) deemed problematic on the basis of other sources' light curves, or 3) cadences that deviated by more than 5 times the reported error from both cadences before and after.  These 444 bad cadences (in red in Figure 2) were eliminated, leaving a total of $55,209$ good cadences.

\subsection{Moir\'{e} pattern drift noise}

\kepler\ data can also suffer from Moir\'{e} pattern drift (MPD) noise, which arises from crosstalk between the four fine guidance sensors and the 84 science channel readouts \citep{kj10}.  
Both \cite{Wehrle13} and \cite{Revalski14} have noted that MPD noise can be a serious problem for \kepler\ AGN variability studies, adding to the apparent source variablity or even mimicking variability in a non-variable source.

There is no known procedure to flag or mitigate MPD errors, although the \kepler\ project is investigating the phenomenon \citep{Clarke14}.  
Table 13 of the \kepler\ Instrument Handbook\footnote{See {\tt https://archive.stsci.edu/kepler/manuals/KSCI-19033-001.pdf}} lists as problematic only one module.output used to observe \zw: 14.4, during Season 2.
The possibility that MPD noise is affecting these data is discussed in Section 3.3.

\section{PSD estimation}
\label{sect:psd}

We performed two power spectral analyses of these data.  
The first uses standard Fourier methods to directly estimate the PSD, the second uses a continuous random process model fitted in the time domain to infer the PSD.

\subsection{Periodogram analysis}

The first approach used the standard periodogram of the full dataset to directly estimate the PSD. 
Gaps within each segment were filled by interpolation (using the LOWESS method; \citet{Cleveland81}) to give one evenly sampled light curve with sampling of $\approx 29.4$ min. 
The light curve was end-matched \citep{Fougere85} to suppress the effects of spectral ``leakage'' \citep{Uttley02}. 
This involves subtracting a linear term such that the mean fluxes for the first and last $20$ data points are the same
The resulting periodogram is shown in Figure 3.

At high frequencies the power spectrum flattens, as expected from independent (white) flux measurement errors, but the observed level is higher than expected given the pipeline errors by a factor $\approx 1.57$, suggesting the flux measurement errors are $25$\% larger than the pipeline errors. 
Further, the PSD rises slowly from $10$ day$^{-1}$ down to $1$ day$^{-1}$, which could be explained in terms of some degree of correlation in the measurement errors, perhaps resulting from MPD. 
At lower frequencies the periodogram rises steeply with decreasing frequency and shows a bend around a timescale of $\sim 5$~days.

We fitted simple models to the periodogram by maximizing likelihood to estimate of the model parameters \citep[see e.g.][]{Vaughan10} using XSPEC 12.8.1 \citep{Arnaud96} with the {\it Whittle} statistic. 
A model comprising a simple power law plus a constant gave a poor match to the data, with power law slope $-3.15$ and fit statistic $D = -2 \log(likelihood) = -270489.0$ using $3$ free parameters. 
Including an additional power law to model the excess power at $\sim 1$ day$^{-1}$ improved the fit by $\Delta D = 157.9$ using five free parameters. 
Including a bend, in the steep power law, to a flatter slope at low frequencies (using a simple bending power law as in \citealt{Edelson13, Gonzales12}) improved the fit by a further $\Delta D = 231.3$,  and has a total of seven free parameters. 
The best-fitting parameters of this model are as follows: power law slopes of $-2.00 \pm 0.12$ at low frequencies and $-4.51\pm 0.20$ at high frequencies, with a bend at $f_b = 0.18 \pm 0.03$ day$^{-1}$ ($\sim 5.6$ day timescale). 
The additional (unbending) power law had a slope $-1.28\pm0.13$ and contributes significantly only around frequencies $\sim 1$ day$^{-1}$. 
Replacing the simple bending power law with a ``Nuker'' law \citep[equation 1 of][]{Lauer05}, which includes an extra parameter to adjust the sharpness of the bend between power law slopes, did not significantly improve the fit ($\Delta D = 0.8$ improvement for one additional free parameter).

\subsection{CARMA-based PSD analysis}

The second analysis utilized the continuous-time autoregressive moving average (CARMA) modeling technique of \citet{Kelly14}.  
This method naturally handles data gaps, measuring the power spectrum down to the lowest frequencies available.  
The CARMA modeling technique assumes that the light curve is a Gaussian process and that the power spectrum can be approximated as a mixture of Lorentzian functions. 
For computational purposes, we reduced the sampling by binning on $2.5$ hour intervals. 
We considered CARMA$(p,p-1)$ models and used the Deviance Information Criterion (DIC, \citealt{Spiegelhalter02}) to choose the value of $p=5$; higher values of $p$ produced worse DIC and did not lead to significantly different power spectra with the exception of higher uncertainty at the low and high frequency ends. 

The inferred power spectrum (Figure 4a) shows evidence for a bending power-law shape, and a flattening toward the highest frequencies. 
The apparent kink in the PSD estimate is an artifact of the use of Lorentzians in the CARMA-PSD process.
By analogy with a bending power-law model, we can quantify the effective power-law slopes of the power spectrum above and below the bend frequency as $d \log P(f) / d \log f$, and the effective bend frequency as the argumentative maximum of $|d^2 \log P(f) / (d \log f)^2|$. 
Based on the CARMA model we infer the effective bend frequency to be located $0.25 \pm 0.01$ day$^{-1}$ ($\sim 4$ day timescale), and  effective power-law slopes of $-1.99 \pm 0.01$ and $-3.65 \pm 0.07$ at frequencies of $0.01$ day$^{-1}$ and $0.6$ day$^{-1}$.

Figure 4b shows the same data with the y-axis multiplied by frequency so equal power per logarithmic frequency interval would be a flat line (e.g. \citealt{Psaltis99}, \citealt{Uttley02}).
Note the continued rise to the longest timescales sampled by \kepler, indicating that the PSD must undergo second flattening because otherwise the total variability power would diverge.

\subsection{Excess high frequency power}

One unusual aspect of these PSD estimates is the apparent excess power at high temporal frequencies.
In particular the upward bend around $\sim 1$~day timescales is unlikely to be intrinsic to the source, since both X-ray binary and AGN $power*frequency$ plots are generally downward-bending or flat at both high and low frequencies, consistent with finite total power (\citealt{Psaltis99}, \citealt{McHardy04}).
No such feature has ever been seen in the X-rays for AGN (e.g., \citealt{Markowitz03}), or, to our knowledge, at any wavelength in any astrophysical source.
In this section we investigate the possibility that this is due to a poorly-studied \kepler\ instrumental effect: MPD noise.

As mentioned earlier, the Kepler Instrument Handbook listed only one of the modules used to observe \zw\ as problematic: the one used for Season 2 data collection.
To investigate this we eliminated Q17 (which is much shorter than all others), then segregated the four quarters in Season 2 (Q4/8/12/16) from the other three seasons (nine quarters in total).
We truncated each quarter's data to the first 67.0 days, the length of the next shortest quarter (Q8).
Then we interpolated across missing cadences in each dataset as before, measured the periodogram, and averaged the Season 2 quarters to produce one PSD estimate (the ``MPD-flagged'' estimate, shown in red in the top panel of Figure 5) and the non-Season 2 quarters to produce another (``MPD-unflagged,'' in blue).  
This assures both PSDs cover exactly the same range of temporal frequencies.

Note that the fits to the MPD-flagged PSD show that both the unbroken $\alpha \sim -1$ power-law and the Poisson noise term are higher for the MPD-flagged PSD, although both PSDs show an excess above the Poisson level expected solely on the basis of the quoted Kepler errors (see \citealt{Vaughan03} for details).
This is clear evidence that there is additional variance in the data on short timescales that cannot be explained solely with the quoted \kepler\ errors.

The ratio of powers (MPD-flagged divided by MPD-unflagged) is shown in the bottom panel of Figure 5.
Note that the MPD-flagged PSD estimate significantly exceeds that of the MPD-unflagged PSD at all temporal frequencies above 1.3~day$^{-1}$ (timescales below $\sim$0.75~day).
This suggests, but by no means proves, that MPD noise could be responsible for the excess high frequency noise, as both the PSD levels and the fits to the two highest-frequency components are larger for the MPD-flagged data than for the MPD-unflagged data.
It also suggests that even the MPD-unflagged quarters suffer from some effects of MPD, because the high-frequency fits and PSD data lie significantly above the levels expected from pure Poisson noise.
However because MPD noise is a detector-wide phenomenon that is poorly suited to study with small ``postage-stamp'' downloads, this indication must be considered tentative until the \kepler\ project completes a systematic detector-wide study of MPD noise.

\subsection{Emission line variability}

Although \cite{Barth11} find that the optical H$\beta$ emission lags the optical continuum by $\sim$4~days, it is not likely that the $\sim 5$~day PSD break is due to emission line variability.  
Examination of spectra taken in that campaign indicates roughly 2-4\% of the total flux in this small ($\sim2x4$~arcsecond) slit due to emission lines.
Further, optical imaging suggests the underlying galaxy contributes no more than $\sim$60\% of the light in this aperture (Q. Wang, priv. comm.), so no more than 40\% is nonstellar.
This means that the emission lines contribute no more than 10\% of the nonstellar Kepler-band flux from the AGN.
(Galactic starlight does not contribute to the PSD because it is not variable.)

In AGN, the fractional continuum variability amplitude is always greater or similar to that of the nearby emission lines. 
Power is the square of amplitude, so the $<10$\% contribution of the lines to the total variability amplitude means that a negligible fraction ($<$1\%) of the variability power can be due to the lines.
Thus we conclude that the PSD features observed in this source are not due to the emission lines, but instead must be due to the nuclear continuum variability.

\section{Discussion}
\label{sect:disco}

The main finding of this paper is that the \kepler\ PSD of \zw\ shows smooth, power law-like shape with a bend from an index of $\sim -2$ at low frequencies to $\sim -4$ at high frequencies at a temporal frequency corresponding to a $\sim 5$ day timescale.
Previous PSD analyses of \zw\ either reported no evidence of a break \citep{Mushotzky11, Kelly14} or a possible break on a $\sim 90$~day timescale \citep{Carini12}.
For the $\sim -2$ slope measured in the current paper, the \citep{Carini12} analysis does not require a break at $\sim 90$~day timescales ($<$60\% likelihood). 
Further the \cite{Carini12} binning limits the high frequency sampling such that it would be difficult to find a high frequency break similar to that reported in this paper.
We thus conclude that we do not have a fundamental disagreement with any previous work. 

\subsection{Accretion disk size and time scales}

In this section we calculate disk sizes implied by the measured $\sim 5$ day timescale under different assumptions about the physical processes that may be responsible for the observed variability.
The optical emission from AGN is thought to arise in an accretion disk so we utilize the standard $\alpha$-disk scaling formulae from \cite{King08}.

Ground-based emission line monitoring of \zw\ \citep{Barth11} yielded a $\sim 4$ day continuum-H$\beta$ lag, which implies a black hole mass of $M_{BH} \sim 10^7$ M$_\odot$ and Schwarzschild radius of $r_S =  2GM/c^2 \approx 3 \times 10^{12}~cm \approx 100$ lt-sec.
\cite{Barth11} also estimated the bolometric luminosity $L_{bol} = 9 \times \lambda L_{\lambda}(5100{\rm A}) = 10^{43.8}$~erg/s, corresponding to $L/L_{Edd} \sim 0.05$.

We consider a range of assumed accretion disk sizes because AGN accretion disk sizes are not currently well-constrained.
First, \cite{Horne14} used contemporaneous \kepler, {\it Swift} and {\it Suzaku} monitoring of \zw\ to estimate the X-ray/optical delay map and thus infer the size of the optically-emitting accretion disk.
That analysis yielded a mean X-ray/optical lag of $\sim 1.7$~days, corresponding to a disk radius of $R \sim 1500 r_S$, albeit with large uncertainties.
Second, we note that standard $\alpha$-disk models (e.g., \citealt{Shakura73}) yield significantly smaller sizes,  $R \sim 100 r_S$.
In order to capture the large uncertainty in this important but poorly-constrained parameter, we consider the implications of the observed $\sim 5$ day break timescale in light of two assumed order-of-magnitude disk size scales: 100 $r_S$ and 1000 $r_S$.
Recent observations of microlensing in AGN (e.g., \citealt{Jimenez14}) typically yield disk size estimates between these values.

\subsubsection{Reprocessing}

``Reprocessing'' models posit that the observed optical variations are due to irradiation of the disk by the central X-ray source (a corona or jet).
The disk would act as a low-pass spatial filter so that the long timescale X-ray variations would be reproduced in the optical while timescales shorter than the light-crossing time would be smoothed out.
This model has been successful in reproducing interband optical correlations in AGN (e.g., \citealt{Cackett07}) but there is often insufficient X-ray luminosity to power the total disk luminosity.

The light-crossing time for a $100-1000 r_S$ emitting region would be $10^4-10^5$~sec or $0.12-1.2$~day.
While the smaller estimate is too small to be consistent with the observed bend timescale, the larger size is marginally consistent, especially given that the light-crossing time is more likely to be associated with the source diameter, not its radius.
Further even in the case of the smaller size estimate, reprocessing would remain viable if the (currently undetermined) PSD of the driving X-ray light curve already had a $\sim 5$~day bend similar to that observed in the optical.
Alternatively it could also be that a small fraction of the optical variability was due to reprocessing, while the bulk due to ``intrinsic'' processes discussed below, in which case the signature would not be visible in the optical PSD.

\subsubsection{Dynamical processes}

For a $10^7 M_\odot$ black hole, the dynamical (orbital) timescale is $ t_{dyn} \sim 140 (R/r_S)^{3/2} $~sec (equation 11, \citealt{King08}).
Optical emission distances of $ R = 100 - 1000 r_S $ yield effective timescales of $t_{dyn} \sim 1.6 - 50 $~days.
This range encompasses the observed $ \sim 5 $~day beak timescale.
It is thus tempting to associate the bend with the dynamical timescale at the radius at which the optical photons are produced.
If this is indeed the proper scaling then we predict that for the other objects observed by \kepler\ that the bend frequency will scale weaker than linearly with the mass and weakly with the Eddington ratio of the sources.

\subsubsection{Thermal processes}

The thermal and viscous timescales are considerably longer \citep{King08}: 
$ t_{dyn} \sim \alpha t_{th} \sim \alpha (H/R)^2 t_{visc} $, where $\alpha$ is the \cite{Shakura73} viscosity parameter and $(H/R)$ is the ratio of disk height to radius.
This assures that $t_{lc} < t_{dyn} < t_{th} < t_{visc}$, since both $\alpha$ and $H/R$ must be less than one. 
For $\alpha \sim 0.1$, an emission distances $ R = 100 - 1000 r_S $ yields $ t_{th} \sim 16 - 500 $~days.
The lower end of this range could be considered a marginally acceptable match to the observed $\sim 5$~day timescale given the uncertainties.

\subsubsection{Viscous processes}

Emission distances of $ R = 100 - 1000 r_S $ yield $ t_{visc} \sim 4 - 140 $~yr for $H/R \sim \alpha \sim 0.1$.
These are not consistent with the observed $ \sim 5 $~day timescale.
This might be considered to be evidence against the propagating fluctuation model \citep{Lyubarskii97, King04, Arevalo06}, which posits that the variations are generated internally by random variations in viscosity (and therefore local accretion rate) over a wide range of spatial scales in the accretion flow. 
Since there must be a second, currently unobserved bend in the PSD at longer timescales than probed by the \kepler\ data in order for the total variability power to be finite (see Figure 4b) this feature may turn out to be associated with viscous or thermal timescales.
The true origin of the observed bend frequency awaits the development of detailed accretion disk models \citep{Schnittman13} capable of self consistently calculating the emitted radiation. 

If the reprocessing model explains the origin of the rapid optical variations, these should be correlated with the X-rays, but delayed and smoothed. 
By contrast if the disk is varying due to intrinsic dynamical/thermal fluctuations, the optical variations may be largely independent of the X-ray variations, or may even generate delayed modulations in the X-rays. 
These predictions can be tested with coordinated X-ray and optical campaigns that provide sufficient signal/noise and temporal resolution.

\subsection{Caveats}

While this is the first optical AGN PSD to span such a large range of temporal frequencies, it is important to be cautious interpreting these new, relatively untested \kepler\ data. 
As discussed in Section 3.3, this PSD seems to show excess power on timescales $<$1~day.
It is tempting to associate this feature with MPD noise (which also seems to show correlated variability on timescales of order $\sim 1$~day) but this has not been definitively established \citep{kj10}.
Further, other possible systematic errors may be present in ways not yet considered.

Finally a more fundamental limitation may be that the underlying PSDs of AGN are not fully described by the simple multiple power-law models used here and in previous analyses.
The highly variable blazar W2R1926+42 exhibits an even more complex \kepler\ PSD than \zw\ \citep{Edelson13}.
Because \kepler\ PSDs cover an unprecedented range of temporal frequency ($\sim 4$ decades), we should remain skeptical of assigning physical significance to parameters derived from simple model fits.

\section{Conclusions}
\label{sect:concl}

This paper presents the best AGN optical light curve ever measured, showing strong variability on long timescales (Figure 2b) yet appearing remarkably ``smooth'' on timescales shorter than a few days (Figure 2c).
The resulting PSD (Figures~3 and 4) is the first to cover nearly $\sim 4$ decades of temporal frequency.
The PSD shows no evidence of QPO, instead it has a smooth, power law-like shape with a bend from an index of $\sim -2$ at low frequencies to $\sim -4$ at high frequencies at a frequency corresponding to a $\sim 5$ day timescale, and excess power near $\sim 1$ day timescales that may be due to systematic MPD noise generated by \kepler. 
This measured timescale can be plausibly associated with the light-crossing, dynamical or thermal timescales, depending on the assumed size of the optically-emitting accretion disk as well as other unobserved disk parameters.
It cannot be associated with the viscous timescale, although the fact that the PSD continues to rise to the longest timescales measured means that there must be a second low-frequency break, and that feature could be associated with the viscous timescale.

These results should also motivate future work in this area.
First, while \zw\ is the best-observed \kepler\ AGN, there are others with data that are nearly as good, and we will be analyzing them in a future paper to see if they share the same characteristics.
Second, the steep PSD at the longest timescales probed by \kepler\ means that there must be a second, currently unobserved break.
It is important to continue to monitor \kepler\ AGN at lower cadence so we can locate that break.
Third, MPD noise is the result of spurious ``waves'' across large regions of \kepler\ detectors, so it cannot be properly addressed by studies limited to tiny ``postage stamp'' downloads such as this.
The \kepler\ project's close-out plans include a final update to the data processing pipeline to identify and flag data impacted by MPD in all \kepler\ light curve files.
Fourth, \kepler\ data are of such high precision that they are pushing the limits of PSD analysis followed by simple broadband model fits.
It would be useful for theorists develop the stationary conditional probability distribution that describes the evolution of the light curve, instead of only the first and second moments of the distribution, which is what the PSD supplies.

\acknowledgments

The authors appreciate the assistance of Martin Still and Tom Barclay of the \kepler\ GO office in helping us understand the \kepler\ data, and Keith Horne and Qian Wang for sharing their unpublished results on \zw.
We also thank the anonymous referee and the editors of {\it the Astrophysical Journal} for a prompt and helpful refereeing process.
This research utilized the HEASARC, IRSA, NED and MAST data archives and the NASA Astrophysics Data System Bibliographic Service.  
RE acknowledges support from the \kepler\ GO and ADAP programs through NASA grants NNX13AC26G and NNX13AE99G.

{\it Facilities:} \facility{\kepler}.

\begin{deluxetable}{lrrrrr}
\tablecaption{Interquarter offset correction
\label{tab:offsets}}
\tablewidth{0pt}
\tablecolumns{6}
\tablehead{
  \colhead{(1)} & \colhead{(2)} & \colhead{(3)} & \colhead{(4)} & 
  \colhead{(5)} & \colhead{(6)} \\
  \colhead{Gap} & \colhead{Quarters} & \colhead{Seasons} & 
  \colhead{Offset} & \colhead{Correction} & \colhead{Residual} \\
  \colhead{} & \colhead{} & \colhead{} & 
  \colhead{(c/s)} & \colhead{(c/s)} & \colhead{(c/s)} 
}
\startdata						
1	 & 4/5 	 & 2/3 & 1471 & -1547 & -76 \\
5	 & 8/9 	 & 2/3 & 1550 & -1547 &	  4 \\
9	 & 12/13 & 2/3 & 1600 & -1547 &	 54 \\
13 & 16/17 & 2/3 & 1566 & -1547 &	 19 \\
	\hline		
2	 & 5/6   & 3/0 & -870 & 750   & -120 \\
6	 & 9/10  & 3/0 & -702 & 750   &	 48 \\
10 & 13/14 & 3/0 & -677 & 750   &	 72 \\
	\hline		
3	 & 6/7   & 0/1 & -234 &	252   &	 18 \\
7	 & 10/11 & 0/1 & -291 &	252   &	-39 \\
11 & 14/15 & 0/1 & -230 &	252   &	 22 \\
	\hline
4	 & 7/8   & 1/2 & -397 &	481   &	 84 \\
8	 & 11/12 & 1/2 & -485 &	481   &	 -3 \\
12 & 15/16 & 1/2 & -562 &	481   &	-81 \\
  \hline
\multicolumn{5}{r}{Standard deviation of residual} & 63 \\
\enddata
\tablecomments{Column 2 lists the \kepler\ quarter numbers bounding each gap.
Column 3 is the same referenced to the four \kepler\ seasons.
Column 4 gives the measured interquarter offset.
Column 5 gives resulting seasonal correction.
Column 6 gives the residual after correction.
The standard deviation of these residuals is given in the bottom line.}
\end{deluxetable}

\begin{figure}
\includegraphics[angle=0,width=6in]{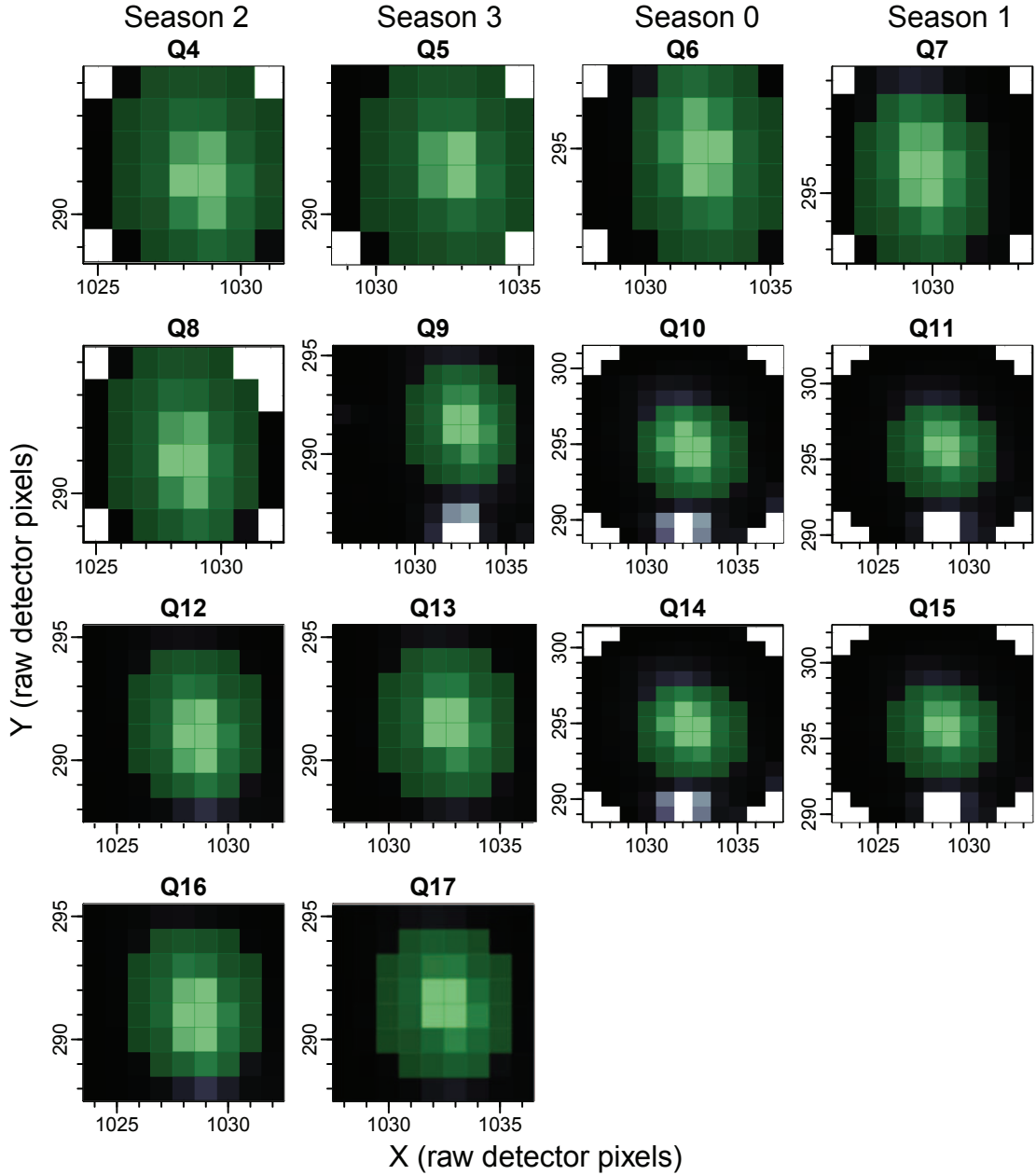} 
\caption{Kepmask output showing downloaded data ``postage stamp'' in greyscale, except occasional missing corner pixels (white).  
Extraction masks (green) are identical within each season and matched as closely as possible between seasons.  
A nearby star is visible e.g. in Q9-11, however its light is not included in the extracted light curve because it lies outside the green extraction mask. 
Only data from pixels within the mask are included in the light curve. 
The 32-pixel masks are the largest possible symmetrical masks given the small Q4/Q5 downloads.  
We find that this makes interquarter repeatability better than the SAP pipeline ``optimal'' masks, which vary in size and location. 
The larger sizes mitigate excursions due to thermally-induced focus changes and to differential velocity aberration \cite{Kinemuchi12}.
\label{fig1}}
\end{figure}

\begin{figure}
 \includegraphics[angle=-90,width=6.5in]{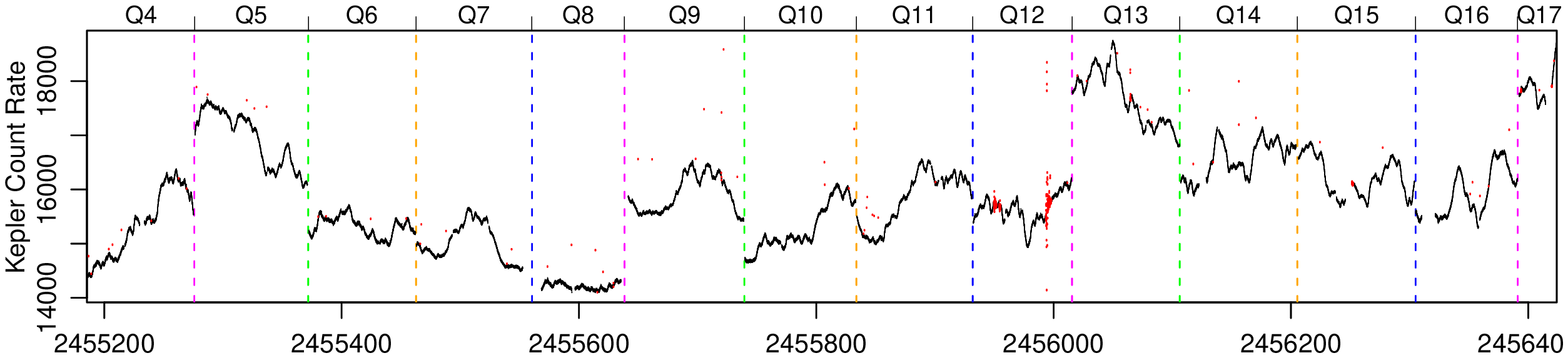} \\
 \includegraphics[angle=-90,width=6.5in]{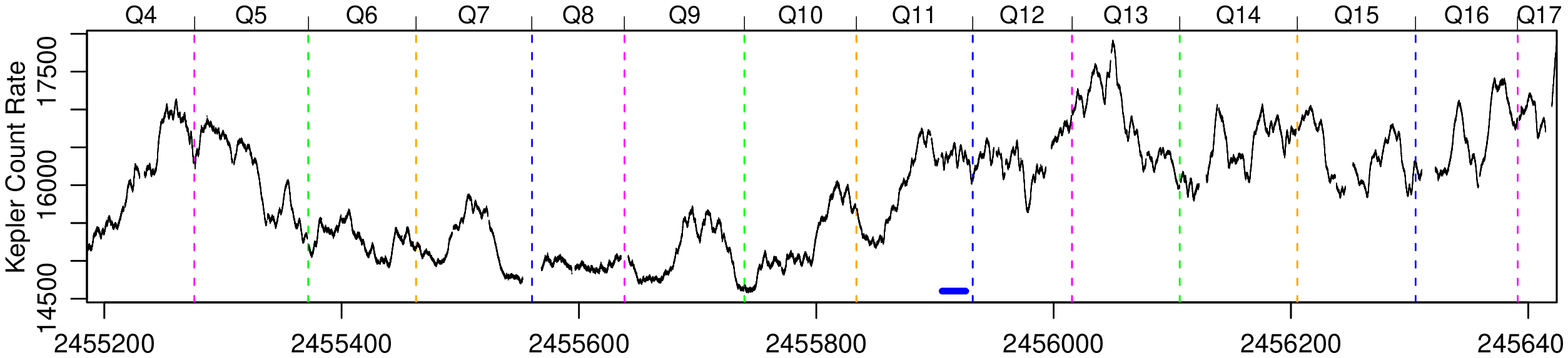} \\
 \includegraphics[angle=-90,width=6.5in]{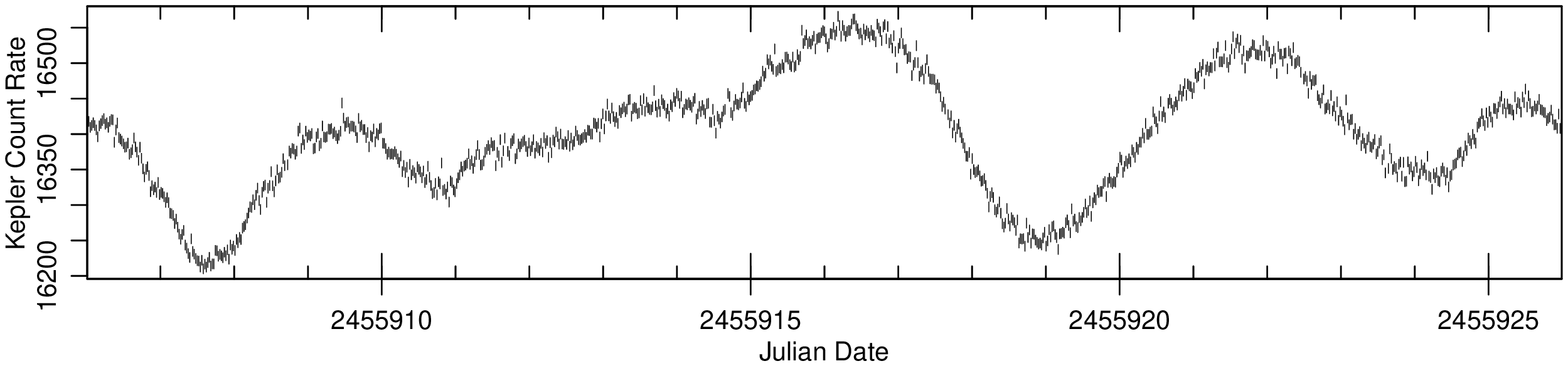}
\caption{2a: Uncorrected light curve.  
Bad data (in red) were eliminated and seasonal jumps were corrected as discussed in text.  
2b: Corrected light curve.  
2c: 20 day (960 cadence) snippet from Q11, showing the quality of the \kepler\ data.  This time range is shown with a horizontal blue line in Figure 2b.
\label{fig2}}
\end{figure}

\begin{figure}
 \includegraphics[angle=-90,width=3in]{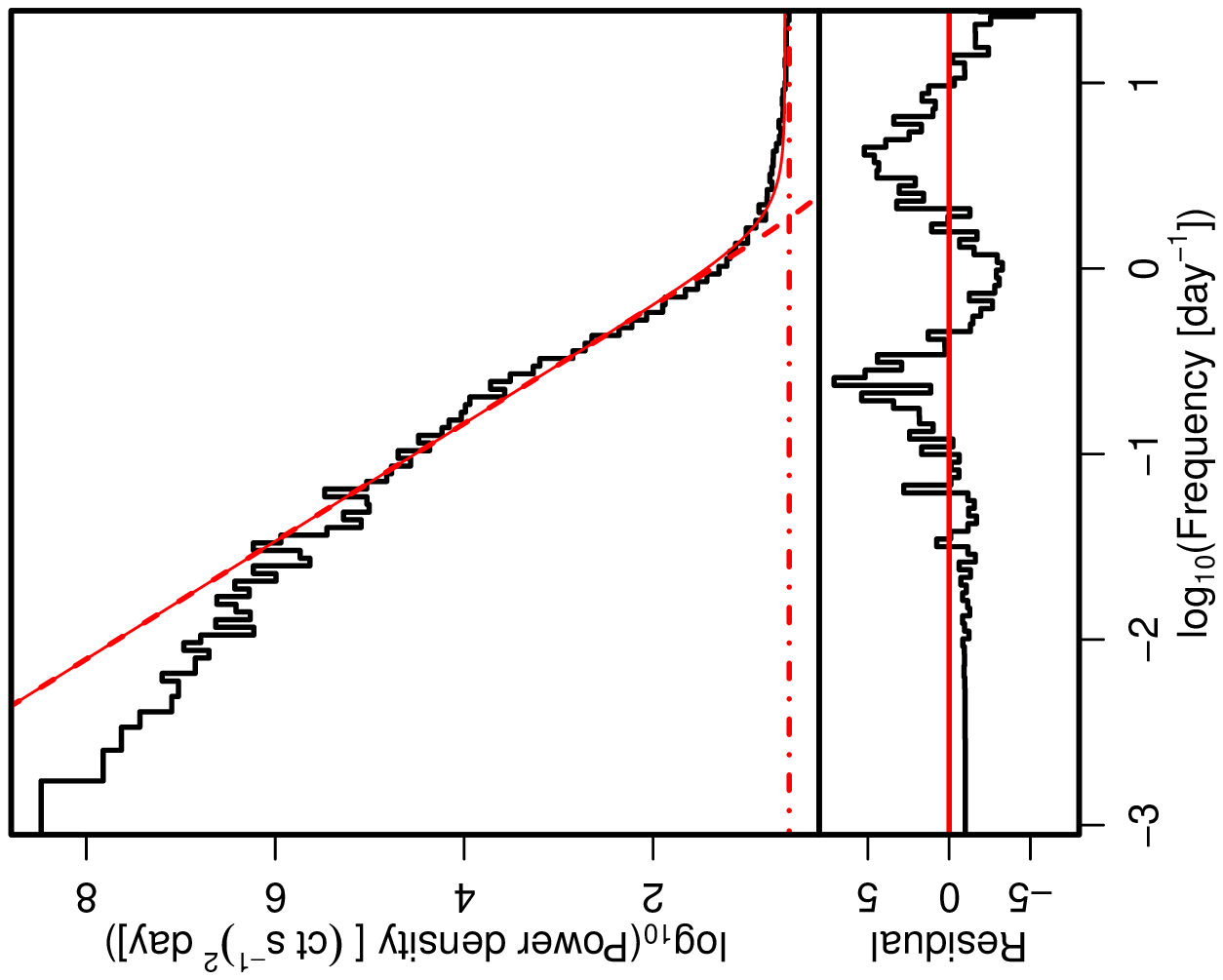} 
 \includegraphics[angle=-90,width=3in]{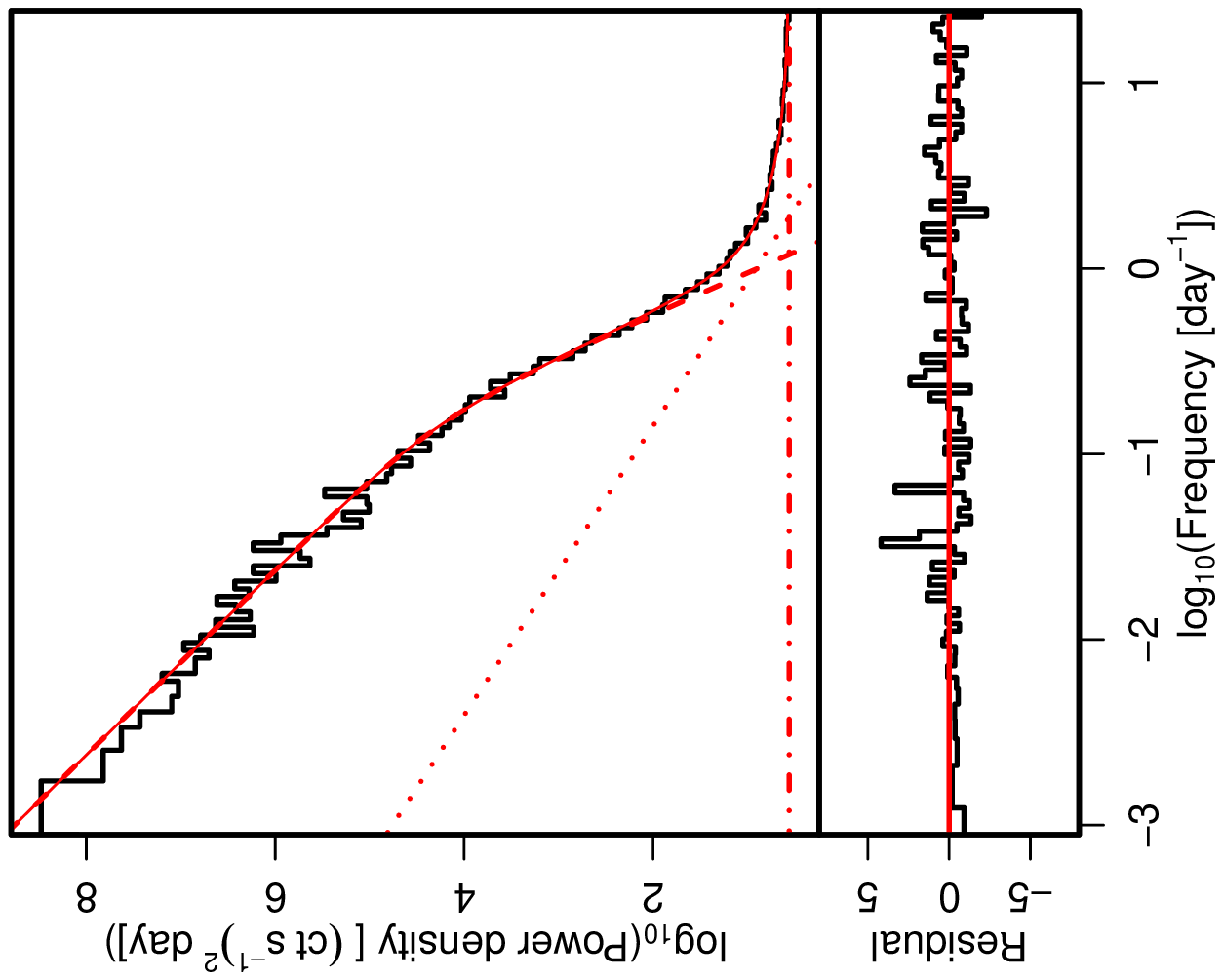} 
\caption{Standard periodogram analysis.
The data (black) have been rebinned for display purposes only, such that the lowest frequency data are not binned, and at higher frequencies data are averaged over bins spanning a factor 1.1 in frequency. 
The fits are shown in solid red at the top and residuals, computed as $(data-model)/(\sqrt{N}*model)$ at the bottom.
3a: A single power-law (dashed line) plus Poisson noise (dot-dash line) model yields a poor fit with large coherent features in the residual plot.
3b: A bending power law (dashed) plus a second power-law (dotted line) plus noise model yields an acceptable fit with smaller and better-distributed residuals.
\label{fig:periodograms}}
\end{figure}

\begin{figure}
 \includegraphics[angle=-90,width=3in]{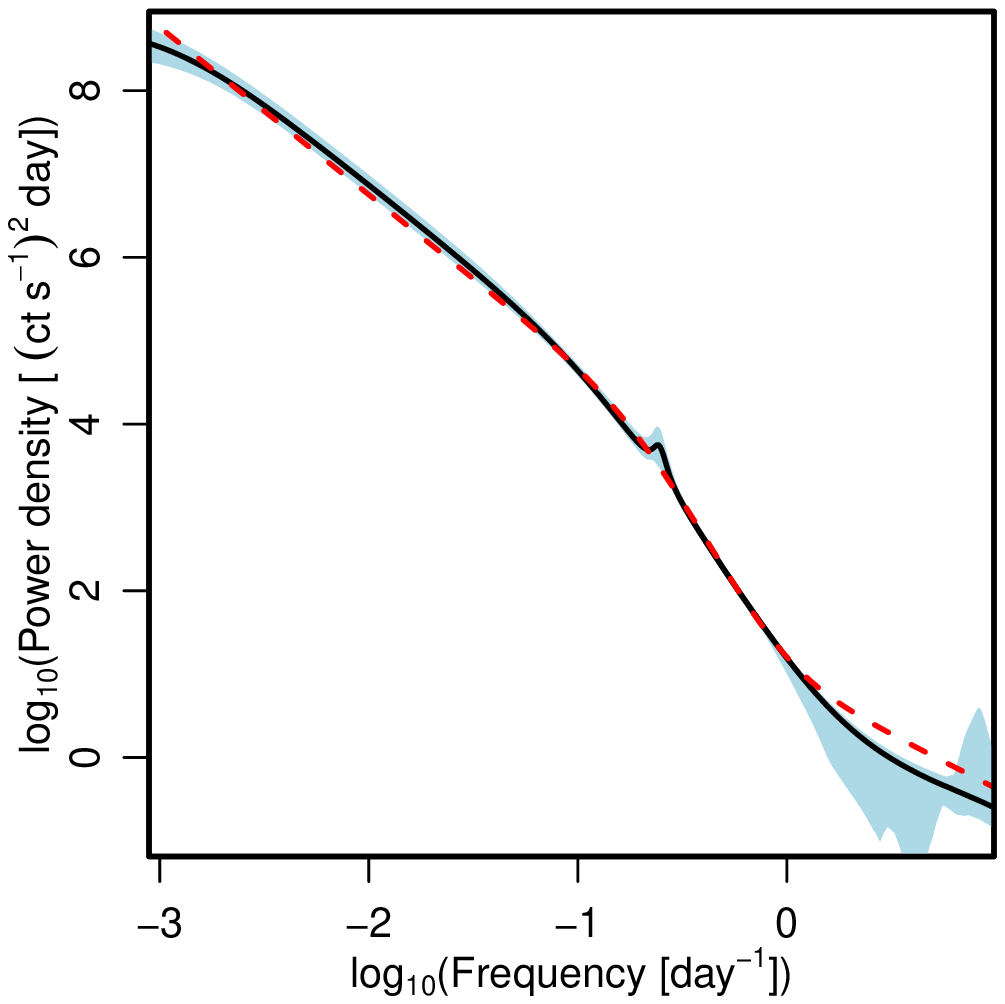} 
 \includegraphics[angle=-90,width=3in]{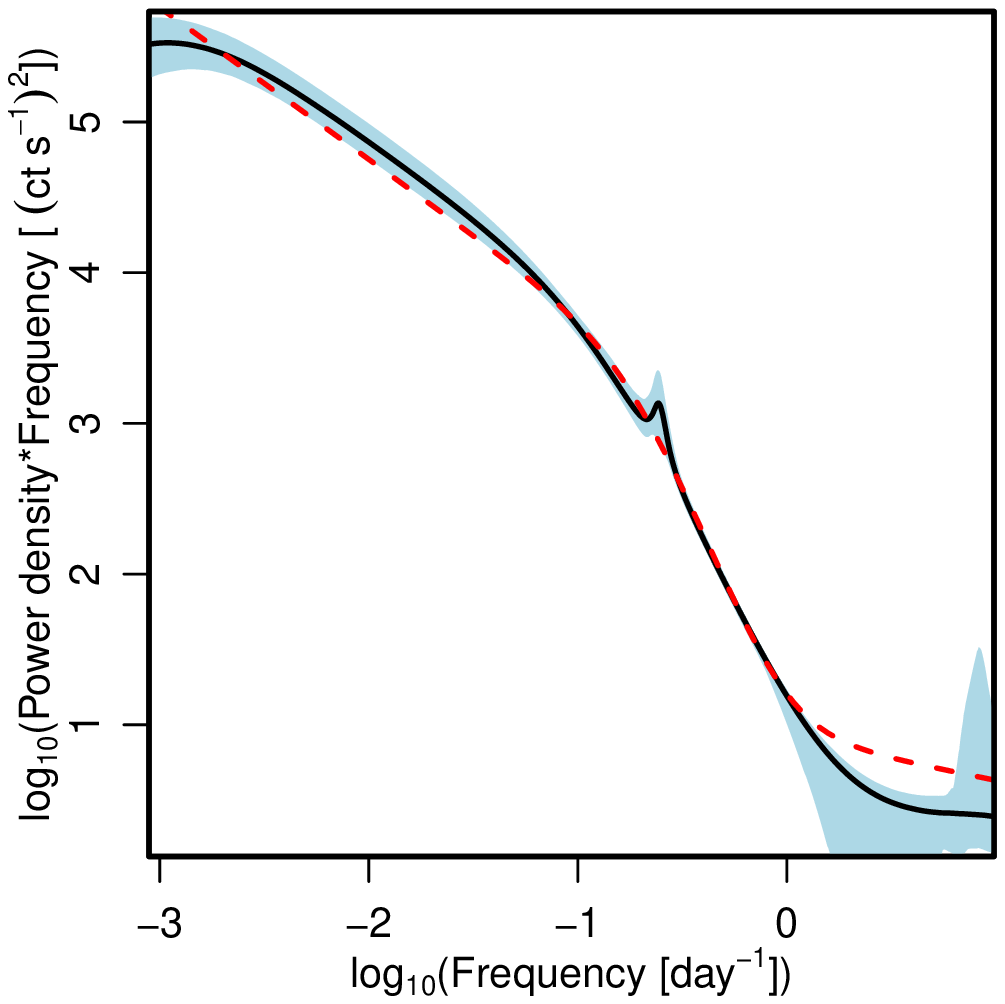} 
\caption{4a: CARMA-based periodogram using the same normalization as Figure 3.  
The best estimate of the PSD is shown as a black line and the shaded blue region shows the $95$\% confidence interval.
Note the good agreement with the Figure 3b fit (dashed red line).
The feature near $\sim 0.3$~day$^{-1}$ is an artifact of the modeling and does not indicate QPO \citep{Kelly14}.
4b: Same data but with the y-axis multiplied by frequency.
\label{fig:carma}}
\end{figure}

\begin{figure}
 \includegraphics[angle=-90,width=3in]{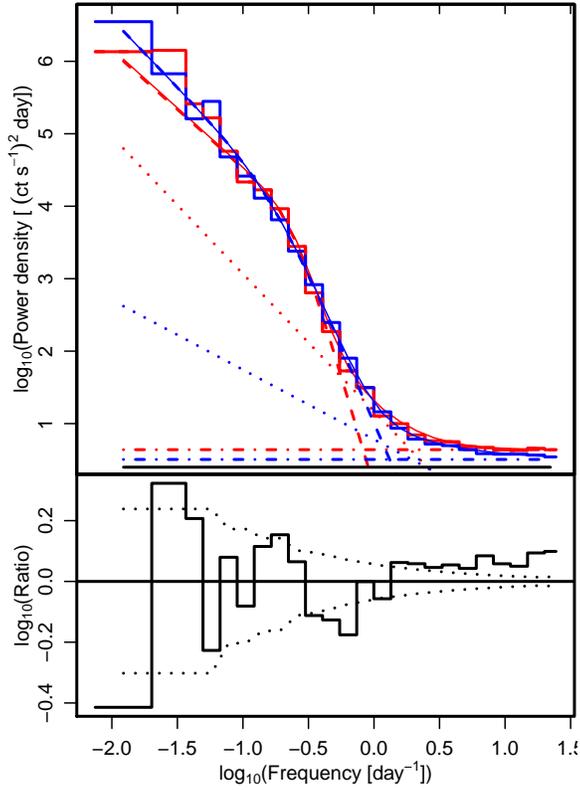} \\
\caption{Top: Averaged PSDs for the MPD-flagged quarters (in red) and MPD-unflagged quarters (blue).  
As in Figure 3, the data are shown as a step plot, full fit as a thin solid line, broken power-law component as a dashed line, second power-law as a dotted line, and Poisson noise as a dot-dash line.
The expected Poisson noise, based on the quoted \kepler\ errors, is shown as a horizontal solid black line.
Note that the MPD-flagged PSD fit at high frequencies shows higher Poisson noise and second power-law than the unflagged PSD fit, and that both stronger Poisson noise than expected from the quoted \kepler\ errors.
Bottom: Logarithm of the ratio of MPD-flagged and unflagged PSDs.
This would be a flat line consistent with zero if MPD was not a source of noise.
Dotted lines show the 1-$\sigma$ confidence interval for a result consistent with zero.
Note that the flagged PSD shows significantly more power than the unflagged PSD at all high frequencies above 1.3~day$^{-1}$.
\label{fig:mpd}}
\end{figure}

\end{document}